\documentclass[aps,prl,twocolumn,showpacs,10pt,superscriptaddress,preprintnumbers,nofootinbib]{revtex4-1}
\usepackage[hyperindex,breaklinks,hyperfootnotes=false,bookmarks=false]%
{hyperref}
\usepackage{graphicx}
\usepackage{amsmath,bm,braket}
\usepackage{color}
\usepackage{multirow}

\allowdisplaybreaks

\def\cO{\mathcal{O}}

\begin{document}

\preprint{MITP/18-060}

\title{$s$-channel Single Top Quark Production and Decay at NNLO in QCD}
\author{Ze Long Liu}
\email{liu@uni-mainz.de}
\affiliation{PRISMA Cluster of Excellence $\&$ Mainz Institute for Theoretical Physics, Johannes Gutenberg University, D-55099 Mainz, Germany}
\author{Jun Gao}
\email{jung49@sjtu.edu.cn}
\affiliation{INPAC, Shanghai Key Laboratory for Particle Physics and Cosmology, School of
	Physics and Astronomy, Shanghai Jiao Tong University, Shanghai 200240, China}

%\pacs{}

\begin{abstract}
\noindent
We report on a fully differential next-to-next-to-leading order (NNLO)
calculation of $s$-channel single top (anti-)quark production with a semi-leptonic
decay at the LHC, neglecting the color correlation between the light and heavy
quark lines and in the narrow width approximation. 
The NNLO corrections can increase the cross section by about 10\% in the low
transverse momentum region of the top quark and reduce scale variation uncertainty. 
In order to compare with experimental results without unfolding procedures, 
we also present theoretical predictions with fiducial cuts, including total cross sections 
and distributions of observables used in the experimental multivariate analysis.
The NNLO corrections are found to be about $-8\%$ for fiducial cross sections. 
\end{abstract}

\pacs{}
\maketitle

\noindent \textbf{Introduction.}
In the Standard Model (SM) of particle physics, the top quark is the heaviest
elementary particle.
The study of top quarks is of great importance for understanding of the nature of
electroweak symmetry breaking and for the fate of the electroweak
vacuum~\cite{Degrassi:2012ry,Alekhin:2012py,Andreassen:2016cvx}.
There are three major modes of electroweak single top quark production at the LHC :
$t$-channel, $s$-channel and $t\,W$ associated production.
The processes are directly sensitive to the 
Cabbibo-Kobayashi-Maskawa (CKM) matrix element $V_{tb}$.
$s$-channel production is of special interest though the cross section
is the smallest.
It is sensitive to new resonances such as $W'$ or
charged Higgs bosons involved in various models beyond the Standard Model (BSM) physics~\cite{cao:2007ea,Tait:2000sh}.
It also serves as an important background process to Higgs studies and
BSM searches~\cite{Chatrchyan:2013zna,Aad:2014xzb,Aad:2014xra,Khachatryan:2016yji}.

$s$-channel single top quark production was first observed by the D0 collaboration
in 2013~\cite{Abazov:2013qka}, and it was confirmed in the combined analysis by
the D0 and CDF collaborations~\cite{CDF:2014uma} at the Fermilab Tevatron.
Recently, it was also measured by the ATLAS and CMS collaborations
at the LHC with 7 and 8 TeV data~\cite{Aad:2015upn,Khachatryan:2016ewo}.
The measurements are expected to enter a precision era with
increasing energy and luminosity of the (HL-)LHC.

To improve the accuracy of theoretical predictions, next-to-leading order
(NLO) QCD corrections for $s$-channel single top quark
production have been calculated with and without considering the
subsequent top quark decay ~\cite{Bordes:1994ki,
	Smith:1996ij,Zhu:2001hw,Harris:2002md,Sullivan:2004ie,Campbell:2004ch,
	Cao:2004ky,Cao:2004ap,Cao:2005pq,Cao:2008af,Campbell:2009gj,Heim:2009ku}.
In Refs.~\cite{Frixione:2005vw,Frixione:2008yi,Alioli:2009je}, the NLO
calculations were also matched to parton shower.
The soft gluon resummations were performed in Refs.~\cite{Kidonakis:2010tc,Zhu:2010mr,Kidonakis:2012rm}.
The NLO QCD correction for $s$-channel production at the LHC is about 35\%,
which is much larger than the estimation from scale variations at leading order (LO).
To control the perturbative uncertainty, it is mandatory to calculate corrections
at higher orders. 

In this Letter, we present a next-to-next-to-leading order (NNLO) QCD
calculation of $s$-channel single top
(anti-)quark production and decay at the LHC using the phase space slicing method.
The inclusive and fully differential cross sections of a stable top (anti-)quark production are
obtained by neglecting the gluon exchange between light and heavy quark lines.
In practice, various kinematic cuts on final states are always
involved in experimental analyses to suppress large backgrounds.
With the known result of top quark decay at NNLO in QCD~\cite{Gao:2012ja},
the fiducial cross sections at the LHC 13 TeV are provided in the narrow
width approximation.
Distributions of various observables within the fiducial volume are
also studied.
These should be helpful for experimental multivariate analyses to
improve the separation between signal and background.

In the following paragraphs we outline the method used in the
calculation and present numerical results on the inclusive and
fiducial cross sections.
Various kinematic distributions are also shown in detail.

\begin{figure}[h!]
	\begin{center}
		\includegraphics[width=0.30\textwidth]{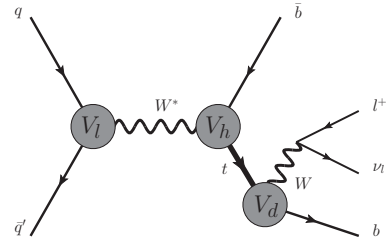}
	\end{center}
	\caption{\label{fig:treefd_2}
		Schematic diagram for $s$-channel single top quark production and decay at hadron
		colliders.}
\end{figure}
\noindent \textbf{Method.}
For $s$-channel single top (anti-)quark production, QCD corrections
can be separated into three categories: corrections associated with
the initial state (light quark line), the final state (heavy quark line)
and gluon exchanges between them.
At NLO, the gluon exchange between the light and heavy quark lines gives no
contribution due to the tracelessness of Gell-Mann matrices.
At NNLO, the color factor of the diagrams with color connection between the two
quark lines are suppressed by $1/N_c^2$ compared with the corrections on the
light or heavy quark lines alone~\cite{Brucherseifer:2014ama}.
Though many efforts have been devoted to calculate two-loop virtual correction in
the color-connected piece of single top quark production~\cite{Assadsolimani:2014oga,Meyer:2016slj},
it is still far from complete.
Here, we treat the corrections for light and heavy quark lines separately,
and neglect color connections between them.
In the narrow width approximation, the top (anti-)quark decay is also included,
of which the NNLO correction has been studied in detail in Ref.~\cite{Gao:2012ja}.
Our strategy can be summarized as in Fig.~\ref{fig:treefd_2}, where
$V_l$, $V_h$ and $V_d$ denote QCD corrections
from the light quark line, heavy quark line and top quark decay, respectively.
All of them are separately gauge invariant and infra red (IR) safe.

To handle the IR divergences, we employ the phase space slicing
method, which has been applied in NNLO QCD calculations of various
processes~\cite{Catani:2007vq,Catani:2009sm,Catani:2011qz,Gao:2012ja,Cascioli:2014yka,Gehrmann:2014fva,Gao:2014eea,Grazzini:2015nwa,Boughezal:2015dva,
Boughezal:2015aha,Boughezal:2015ded,Berger:2016oht,Berger:2017zof,Berger:2016inr}.
A general method named $N$-jettiness subtraction~\cite{Boughezal:2015dva,Gaunt:2015pea} is available to the processes with massless parton in final state. 
For any infrared safe observable $O$, the differential cross section can be expressed as 
\begin{equation}
\begin{aligned}
\frac{d\sigma}{d O}&=\underbrace{\int_0^{\tau_{\rm cut}}d\tau\frac{d\sigma}{d\tau\,dO}}_{\rm unresolved} + \underbrace{\int_{\tau_{\rm cut}}^{\tau_{\rm max}}d\tau\frac{d\sigma}{d\tau\, dO}}_{\rm resolved}\,,
\end{aligned}
\end{equation}
where $\tau$ is a slicing variable.
Below the cutoff $\tau_{\rm cut}$, given $\tau_{\rm cut}$ sufficiently small,
all the radiations are unresolved, i.e. either soft or collinear to the beam or jet axes.
Those contributions can be systematically factorized with soft-collinear effective
theory (SCET)~\cite{Bauer:2000ew,Bauer:2000yr,Bauer:2001ct,Bauer:2001yt,Bauer:2002nz,
Beneke:2002ph} at leading power of $\tau_{\rm cut}$.
Progress has been made to compute the subleading power
corrections~\cite{Moult:2016fqy,Boughezal:2016zws,Moult:2017jsg,Boughezal:2018mvf}. 

For the light quark line, we adopt the $0$-jettiness with
two beam axes as the slicing variable.
For the unresolved part, the factorization formula was derived
in Ref.~\cite{Stewart:2009yx}.
The hard, soft and quark beam functions are available up to
NNLO~\cite{Idilbi:2006dg,Becher:2006mr,Kelley:2011ng,Monni:2011gb,Gaunt:2014xga,Gaunt:2014cfa}.
For the resolved part, the NNLO contribution is equivalent to the NLO cross
section of $pp\to W^* + {\rm jet}$.
The one-loop amplitudes of
$q+\bar{q}'\to W^* + g$ and $q(\bar{q})+g\to W^* + q'(\bar{q}')$ can be
obtained by a non-trivial analytical continuation of the one-loop
amplitudes of $e^+e^-\to q\bar{q}g$~\cite{Garland:2002ak,Gehrmann:2002zr}.
The dipole subtraction~\cite{Catani:1996vz} is employed to deal with
IR divergences at NLO.
By setting the top quark mass $m_t=0$, the NNLO correction has been cross
checked with result from $\texttt{DYNNLO}$~\cite{Catani:2009sm,Catani:2007vq}.

For the heavy quark line, by neglecting the bottom quark mass and clustering all the massless partons in final state into a single jet, the slicing variable
$\tau_{h}$ is defined as $\tau_h={m_J^2}/{Q^2}$, where $m_J$ and $Q$ are
the invariant masses of the jet and off-shell $W$ boson, respectively.
In the limit of $\tau_h\to 0$, all QCD radiation should be soft
or collinear to the bottom quark direction.
The unresolved cross section can be expressed as 
\begin{equation}
\begin{aligned}
\frac{d\sigma_h}{dO}\Bigg|_{\rm unres.}
&=\sum_{q,\bar{q}'} f_{q} \otimes  f_{\bar{q}'} \otimes H_h \otimes S_h \otimes J_q  \\
& +\cO(\tau_{h,{\rm cut}}\ln^k{\tau_{h,{\rm cut}}})\,,
\end{aligned}
\end{equation}
where $f_q$, $H_h$, $S_h$ and $J_q$ are the parton distribution function (PDF),
hard function, soft function and quark jet function, respectively.
The quark jet function is already known up to $\cO(\alpha_s^3)$~\cite{Becher:2006qw,Bruser:2018rad}.
The NNLO soft function can be obtained from Ref.~\cite{Becher:2005pd} by boosting
to the rest frame of the top quark.
The hard function encodes the contribution of virtual corrections,
which only depend on the dimensionless variables $x=(p_b+p_t)^2/m_t^2$ and $L_t=\ln(\mu/m_t)$,
with $\mu$ being the renormalization scale.
$p_t$ and $p_b$ denote the momenta of the top quark and bottom anti-quark, respectively.
In Refs.~\cite{Bonciani:2008wf,Asatrian:2008uk,Beneke:2008ei,Bell:2008ws}, QCD
corrections to the $b\to u$ current was calculated up to $\cO(\alpha_s^2)$ analytically.
The results were expressed in terms of a set of harmonic polylogarithms (HPLs), which
have a well-defined analytical continuation.
Thus, we can use it to derive $H_h(x,m_t^2,\mu)$ by restoring the imaginary part
\begin{equation}
x=\frac{ (p_b+p_t)^2 + i\,\varepsilon}{ m_t^2 - i\,\varepsilon}= \frac{ (p_b+p_t)^2 }{ m_t^2 }+ i\,\varepsilon\,,
\end{equation}
with $\varepsilon$ being an infinitesimal.
As a cross check, we performed analytical continuation of the matching coefficients
in Refs.~\cite{Asatrian:2008uk} and~\cite{Beneke:2008ei} independently and found the same  results.
For the resolved part, there is at least one additional hard jet due to
the phase space constraint  $\tau_h>\tau_{h{,\rm cut}}$.
At NNLO, this contribution can be described by the NLO corrections
to $W^*\to \bar{b} + t +{\rm jet}$.
The one-loop virtual correction can be obtained from Ref.~\cite{Campbell:2005bb} with crossing.
Dipole subtraction \cite{Catani:2002hc} is employed to handle the IR singularities.

\noindent \textbf{Numerical result.}
The relevant parameters used in our numerical calculation are listed as follows.
The top quark and $W$ boson masses are set to $172.5$ GeV and $80.385$ GeV, respectively.
The Fermi constant $G_F$ is chosen as $1.166379\times 10^{-5}~{\rm GeV}^{-2}$.
The CKM matrix elements are set to $\{|V_{ud}|,|V_{us}|,|V_{tb}|\}=\{0.975,\,0.222,\,1\}$.
The default values of the renormalization scale $\mu_R$ and factorization scale
$\mu_F$ are chosen as $\mu_R=\mu_F=m_t$.
The scale uncertainties are calculated by varying $\mu_R$ and $\mu_F$ simultaneously
by a factor of two from the default value.
We use the CT14NNLO PDF set~\cite{Dulat:2015mca} with $\alpha_s(M_Z)=0.118$. 

\begin{figure}[t]
	\begin{center}
		\includegraphics[width=0.43\textwidth]{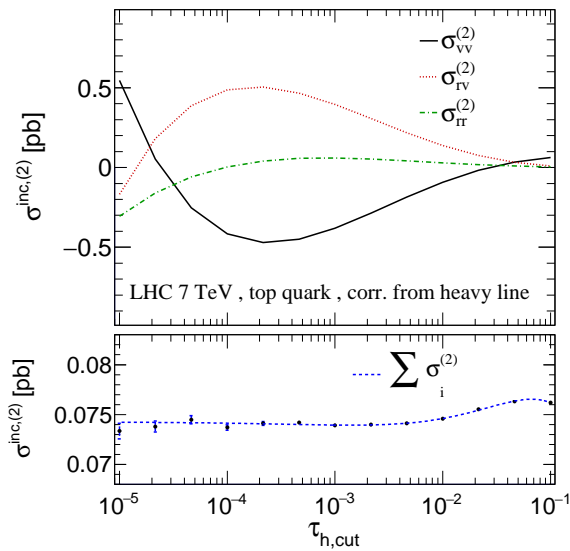}
	\end{center}
	\caption{\label{fig:tauhdep} 
Cutoff dependence of various components of NNLO corrections from 
heavy quark line.
The lower plot shows the sum of $\sigma_{vv}^{(2)}$, $\sigma_{rv}^{(2)}$ and
$\sigma_{rr}^{(2)}$.}
\end{figure}

Fig.~\ref{fig:tauhdep} shows three components of the NNLO corrections
as a function of the cutoff $\tau_{h,{\rm cut}}$ for the heavy quark line.
$\sigma_{vv}^{(2)}$, $\sigma_{rv}^{(2)}$ and $\sigma_{rr}^{(2)}$ denote
the contributions from the two-loop virtual correction, the one-loop
real-virtual correction and the double real correction, respectively.
The sum of them converge smoothly to $0.074~{\rm pb}$ as
$\tau_{h,{\rm cut}}$ approaching 0.
The dependence of the inclusive cross section on $\tau_{h, {\rm cut}}$
are negligible for $\tau_{h, {\rm cut}}$ below $10^{-3}$, as expected.

\begin{table}[t]
	\centering
	\begin{tabular}{l|l|c|c|c} \hline
		\multicolumn{2}{c|}{inclusive }  & LO & NLO & NNLO  \\  [1ex] 
		\hline \hline 
		\multirow{3}{*}{8 TeV} & $\sigma(t)\, {\rm [pb]}$ & $2.498^{  + 0.17\%}_{  -0.74\%}$ &             $3.382^{  + 2.36\%}_{  -1.81\%}$ &             $3.566^{ +  0.95\%}_{  -0.78\%}$   \\  [1ex] 
		& $\sigma(\bar t)\, {\rm [pb]}$ &   $1.418^{  + 0.12\%}_{  -0.73\%}$ &             $1.922^{  + 2.37\%}_{  -1.81\%}$ &             $2.029^{  + 1.07\%}_{  -0.83\%}$  \\  [1ex] 
		& $\sigma(t+\bar t)\, {\rm [pb]}$ & $3.916^{ +  0.15\%}_{  -0.73\%}$ &             $5.304^{ +  2.36\%}_{  -1.81\%}$ &             $5.595^{  + 0.99\%}_{  -0.80\%}$  \\  [1ex] 
		& $\sigma(t)/\sigma(\bar t)$ &    $1.762^{ + 0.04\%}_{  -0.01\%}$ &             $1.760^{  + 0.00\%}_{  -0.02\%}$ &             $1.757^{  + 0.05\%}_{  -0.12\%}$  \\  [1ex]  
		\hline
		\multirow{3}{*}{13 TeV} & $\sigma(t)\, {\rm [pb]}$& $4.775^{ +  2.69\%}_{  -3.50\%}$ &             $6.447^{  + 1.39\%}_{  -0.91\%}$ &             $6.778^{ +  0.76\%}_{  -0.53\%}$  \\  [1ex] 
		& $\sigma(\bar t)\, {\rm [pb]}$ &  $2.998^{ +  2.69\%}_{  -3.55\%}$ &             $4.043^{ +  1.33\%}_{  -0.94\%}$ &             $4.249^{  + 0.69\%}_{  -0.48\%}$  \\  [1ex] 
		& $\sigma(t+\bar t)\, {\rm [pb]}$ & $7.772^{ +  2.69\%}_{  -3.52\%}$ &             $10.49^{  + 1.36\%}_{  -0.92\%}$ &             $11.03^{   +0.74\%}_{  -0.51\%}$\\  [1ex] 
		& $\sigma(t)/\sigma(\bar t)$ &  $1.593^{ +  0.05\%}_{  -0.01\%}$ &             $1.595^{  + 0.06\%}_{   0.03\%}$ &             $1.595^{  + 0.07\%}_{  -0.05\%}$ \\  [1ex] 
		\hline
	\end{tabular}
	\caption{
Inclusive cross section for $s$-channel single top (anti-)quark production at LO,
NLO and NNLO at the LHC  8 and 13 TeV.
The uncertainties refer to the variation by simultaneously changing the factorization and
renormalization scales by a factor of two from their central
value $\mu_F=\mu_R=m_t$.}
\label{tab:inclusive}
\end{table}

In Tab.~\ref{tab:inclusive}, we present the inclusive cross
section of $s$-channel single top (anti-)quark production at the LHC 8 and 13 TeV.
Both of the NLO and NNLO corrections enhance the inclusive cross sections.
The NLO corrections are typically 35\%.
The NNLO corrections are about
7\% in general, indicating a good perturbative convergence.
The scale variations for the LO cross section are quite small due to the
opposite trend of the $u$ and $\bar{d}$ quark PDFs from varying the factorization scale.
The NNLO corrections would be underestimated by the scale variations of the NLO
cross sections.
Nevertheless, the scale variations are largely reduced with the
NNLO corrections.
At NLO, both of the corrections to $V_l$ and $V_h$ are significant.
At NNLO, the corrections to $V_l$ are below 1\%, much smaller compared to the
corrections to $V_h$ and the product of the $\cO(\alpha_s)$ corrections
to $V_l$ and $V_h$, which are more than 2\%.
QCD corrections are similar for top quark and anti-quark production.
The ratio of the two cross sections are  thus stable against QCD corrections,
varying at the per mille level.

\begin{figure}[t]
	\begin{center}
		\includegraphics[width=0.425\textwidth]{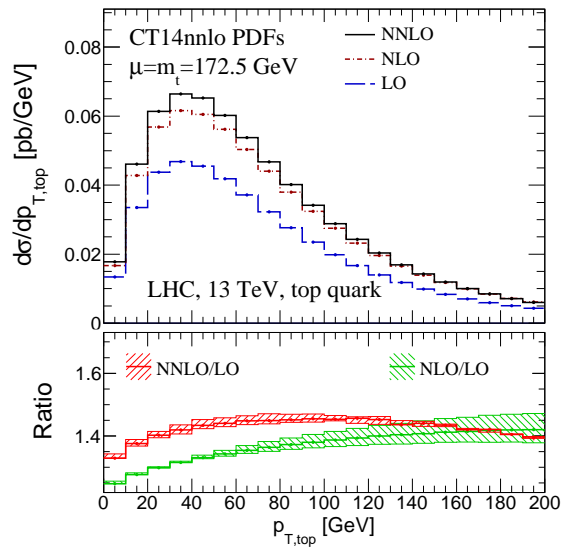}
	\end{center}
	\caption{\label{fig:diffpTtop}
Predicted transverse momentum distribution of the top quark from $s$-channel production at the LHC 13 TeV.}
\end{figure}

Fig.~\ref{fig:diffpTtop} shows the transverse momentum distribution of top
quark at the LHC 13 TeV.
Both the NLO and NNLO corrections are positive and large.
The ratios of NLO to LO cross sections vary from 1.2 to 1.4 over the range
$0 < p_{T,{\rm top}} < 200\,{\rm GeV}$, and the ratios of NNLO to LO cross sections
vary from 1.35 to 1.45 for the same range.
In low $p_{T,{\rm top}}$ region, the NNLO corrections can be as large as 10\%.
There is no overlap between the NLO and NNLO prediction bands in most region,
which again indicates the NNLO corrections would be underestimated by scale variations
at NLO.
The scale variations are greatly reduced going from NLO to NNLO for large
$p_{T,{\rm top}}$ values. 

In experimental analyses, top (anti-)quarks are identified through their decay products e.g.,
semi-leptonic or hadronic decays.
With the advantage of our fully differential calculation, we can study observables
within an experimental fiducial volume.
In the following calculations, we assume top quarks always decay to $bW^+$ and use a branching ratio of 0.1086 for the leptonic decay of the $W$ boson to one family.
Based on the CMS analysis~\cite{Khachatryan:2016ewo}, we choose the following
basic kinematic cuts.
Events with one charged lepton are selected by requiring its transverse mometum
$p_{T,l}>24\,{\rm GeV}$ and pseudorapidity $|\eta|<2.1$.
Jets are clustered with anti-$k_T$ jet algorithm and radius $R=0.5$.
Pre-selection requires jets to have $|\eta|<4.5$ and $p_T>20\,{\rm GeV}$.
Pseudorapidity of bottom quark initiated jets are required to satisfy $|\eta|<2.4$
according to $b$ tagging algorithms~\cite{Chatrchyan:2012jua}.
Single top quark production through $s$-channel is characterized by a final state
composed of one charged lepton, missing energy originating from neutrinos, and
two $b$-tagged jets.
One of the $b$-jets is associated with top-quark production and the other is
from top-quark decay.
We employ the ``$2$-jets $2$-tags'' analysis~\cite{Khachatryan:2016ewo},
which requires exactly two jets, each with transverse momentum greater than 40 GeV,
and both being $b$-tagged.

\begin{table}[t]
	\centering
	\begin{tabular}{l|l|c|c|c} \hline
		\multicolumn{2}{c|}{fiducial [pb]}  & LO & NLO & NNLO  \\  [1ex] 
		\hline \hline 
		\multirow{3}{*}{$t$ quark} & total & $     0.1348_{  -3.4\%}^{   +2.6\%}$ &    $     0.1156_{  -3.0\%}^{   +3.1\%}$ &    $     0.1071_{  -0.8\%}^{   +2.2\%}$   \\   
	    & corr. in pro. &  &  $    -0.0121                        $ &     $    -0.0065                  $         \\ 
		 & corr. in dec. &   & $    -0.0071                        $ &     $    -0.0026                 $        \\ 
		\hline
		\multirow{3}{*}{$\bar t$ quark} & total & $     0.0907_{  -3.4\%}^{  + 2.5\%}$ &    $     0.0745_{  -3.4\%}^{  + 3.6\%}$ &    $     0.0663_{  -1.3\%}^{   +2.5\%}$  \\   
		  & corr. in pro. & & $    -0.0066                        $ &     $    -0.0051                  $        \\ 
		 & corr. in dec. &  & $    -0.0096                        $ &     $    -0.0035                 $       \\ 
		\hline
	\end{tabular}
	\caption{
Total cross section within the fiducial volume at the LHC 13 TeV.
The NLO and NNLO QCD corrections from top quark production and decay are
also listed separately.}
\label{tab:fiducial}
\end{table}

We summarize the total cross sections at LO, NLO and NNLO with the fiducial cuts
at the LHC 13 TeV in Tab.~\ref{tab:fiducial}.
The QCD corrections from production and decay alone are also listed.
In contrast to the inclusive cross sections, both the NLO and NNLO corrections are
negative for fiducial cross sections.
The NLO and NNLO corrections are about $-16\%$ and  $-8\%$, respectively.
QCD corrections from decay are comparable to those from production,
especially for top anti-quark.
The scale variations are reduced with NNLO corrections.

Next, we show distributions of two observables that are key inputs to the experimental
multivariate analysis.
\begin{figure}
	\begin{center}
		\includegraphics[width=0.435\textwidth]{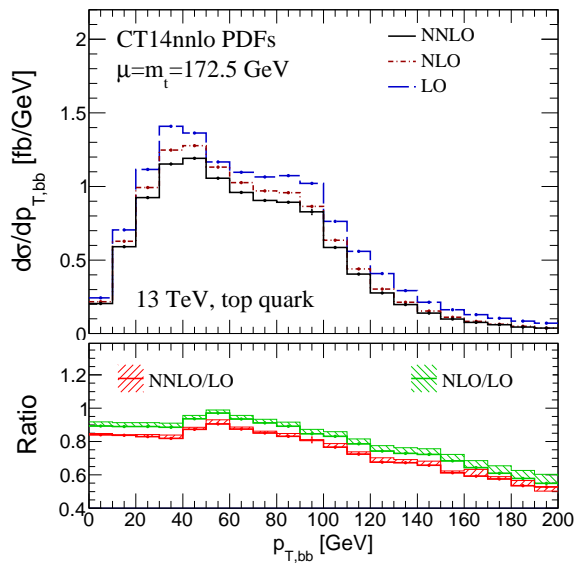}
	\end{center}
	\caption{\label{fig:fid13pTbbtop}
Transverse momentum distribution of	the two $b$-jet system from the $s$-channel
single top quark production and decay at the LHC 13 TeV with fiducial cuts. }
\end{figure}
Fig.~\ref{fig:fid13pTbbtop} presents the transverse  momentum
distribution of the two $b$-jet system in $s$-channel single top quark production and decay.
The NNLO correction to the distribution is about $-10\%$ over the range
$0 < p_{T,{bb}} < 200\,{\rm GeV}$.
There is an obvious gap between the NLO and NNLO prediction bands.
The scale uncertainties are reduced by NNLO corrections especially in large $p_{T,bb}$ region.
\begin{figure}
	\begin{center}
		\includegraphics[width=0.435\textwidth]{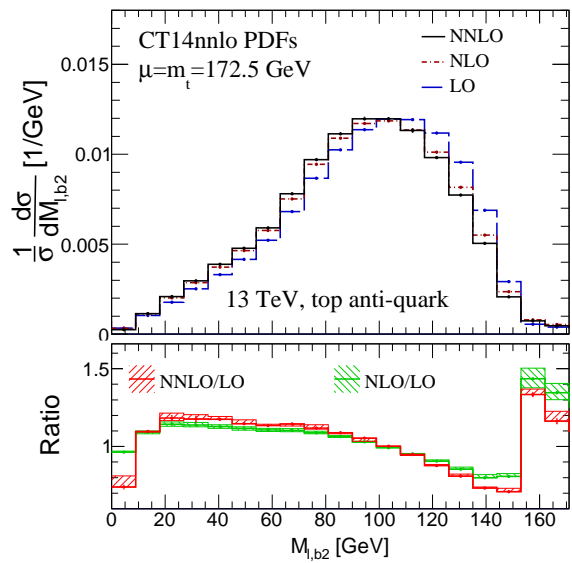}
	\end{center}
	\caption{\label{fig:fid13Mlb2top}
Normalized distribution of invariant mass of the
system composed of the charged lepton and the subleading $b$ jet in $p_T$,
for top anti-quark production and decay at the LHC 13 TeV with fiducial cuts.}
\end{figure}
Fig.~\ref{fig:fid13Mlb2top} presents the normalized distribution of the invariant
mass of the system composed of the charged lepton and the subleading $b$ jet in $p_T$
in $s$-channel top anti-quark production and decay.
The distribution of $M_{l,b_2}$ has an endpoint around the top quark mass, as expected.
The peak of the distribution is shifted to lower masses by higher order corrections.
The normalized distribution show little dependence on the scale choices.
The ratios of NLO and NNLO cross sections to LO ones grow rapidly when $M_{l,b_2}$
increases above $160\,{\rm GeV}$, which is close to the top quark mass threshold.

\noindent \textbf{Conclusions.}
We have presented a first NNLO QCD calculation of $s$-channel single
top (anti-)quark production and decay at the LHC neglecting certain
subleading color contributions.
The top (anti-)quark spin correlation is preserved in the narrow width approximation.
By considering NNLO corrections, the inclusive cross sections are 
enhanced by about 7\% in general.
The increase of cross sections at low transverse momentum of the top quark can
reach above 10\%.
Furthermore, the NNLO corrections to the total fiducial cross section are
about $-8\%$, in contrast to the inclusive case.
The scale variations are reduced in general for both inclusive and fiducial
cross sections.
We found scale variations at NLO always underestimate the true NNLO corrections.
The NNLO corrections are also significant for various kinematic distributions,
including the shapes.
Our results can be used to improve the measurement of cross sections of
$s$-channel single top quark production, extraction of the top quark electroweak
coupling and also the measurement of the top quark mass~\cite{Alekhin:2016jjz}. 

\bigskip	
\begin{acknowledgments}
We thank Hua Xing Zhu for useful communications. We are grateful to Maximilian Stahlhofen and Felix Yu for carefully reading this manuscript and precious comments.
We thank Johannes Gutenberg University Mainz for the use of the High Performance Computing facility Mogon.  
JG's work is sponsored by the Shanghai Pujiang Program. 
Z.L.L is supported by the Cluster of Excellence
{\em Precision Physics, Fundamental Interactions and Structure of Matter\/} (PRISMA -- EXC 1098) at JGU Mainz. 
\\
\end{acknowledgments}
%% Bibliography
\bibliographystyle{apsrev4-1}
\bibliography{s_top_short}

\end{document}